\documentclass[a4paper]{lipics-v2021}

\hideLIPIcs
\nolinenumbers

\pdfoutput=1 

\title{An Infinitary Lambda Calculus with Global Trace Condition (Extended Abstract)}

\author
    {Stefano Berardi}
    {Computer Science Department, Turin University, Torino, Italy}
    {} 
    {0000-0001-5427-0020} 
    {} 

\author
    {Ugo de' Liguoro}
    {Computer Science Department, Turin University, Torino, Italy}
    {} 
    {0000-0003-4609-2783} 
    {} 

\author
    {Daisuke Kimura}
    {Department of Information Science, Toho University, Japan}
    {} 
    {0000-0001-6287-692X} 
    {} 

\author
    {Daniel Osorio-Valencia}
    {Computer Science Department, Turin University, Torino, Italy}
    {} 
    {0009-0002-4820-1867} 
    {} 
\authorrunning{S.\,Berardi et al.}

\Copyright{Stefano\,Berardi and Ugo\,de'\,Liguoro and Daisuke\,Kimura and Daniel Osorio}



\category{} 

\relatedversion{} 

\nolinenumbers 

\EventEditors{John Q. Open and Joan R. Access}
\EventNoEds{2}
\EventLongTitle{42nd Conference on Very Important Topics (CVIT 2016)}
\EventShortTitle{CVIT 2016}
\EventAcronym{CVIT}
\EventYear{2016}
\EventDate{December 24--27, 2016}
\EventLocation{Little Whinging, United Kingdom}
\EventLogo{}
\SeriesVolume{42}
\ArticleNo{23}

\usepackage{ dsfont }
\usepackage{graphicx}
\usepackage[all]{xy}
\usepackage{amssymb,amsmath,graphicx,color}
\usepackage{mymath,latexsym}
\usepackage{upgreek}
\usepackage{ebproof}
\usepackage{xcolor}
\usepackage[pdftex,outline]{contour}
\usepackage{mdframed}
\usepackage{tikz} 
\usetikzlibrary{arrows.meta} 
\usepackage{marvosym}
\usepackage{proofzilla}
\newemptyvertex{node}{}
\defedgetype{flow}{draw=blue}{}
\defedgetype{dirflow}{-Latex,draw=blue}{}
\defedgetype{flowtwo}{draw=red}{}
\defedgetype{dirflowtwo}{-Latex,draw=red}{}
\defedgetype{dirflowblue}{-Latex,draw=blue}{}
\defedgetype{dirflowred}{-Latex,draw=red}{}
\definecolor{black}{gray}{0}
\defedgetype{flowblack}{-Latex,draw=black}{}

\begin{document}

\maketitle

\begin{abstract}
We consider an extension of the infinitary lambda calculus by Kennaway et al., with zero, successor, and conditional, and a type system akin to G\"odel's system T. For terms that can be typed in this system, we define the Global Trace Condition (GTC), adapting the concept from Brotherston and Simpson's Cyclic Proofs, and show that any infinite reduction of a well-typed term satisfying the GTC is strongly convergent. As a corollary, we obtain the proof that any closed term of type Nat reduces to some numeral through any reduction by levels. 

We argue that the Church-Rosser in the limit holds for our calculus and, when restricted to regular terms, the calculus defines exactly the total functions defined in Das's Cyclic System T (an infinitary version of System T without $\lambda$), and hence in G\"odel's System T.
\end{abstract}



\def\Expand{{\rm expand}}


\def\RCLKIDomega{{{\bf RCLKID}^\omega}}
\def\Ren{{\rm Ren}}
\def\Terms{{\rm Terms}}
\def\IE{{\rm IE}}
\def\Cut{{\rm Cut}}
\def\Rcut{{\rm Rcut}}


\def\Co{{\rm co}}
\def\Axiom{{\rm Axiom}}
\def\Wk{{\rm Wk}}
\def\Cut{{\rm Cut}}
\def\BS{{\rm BS}}
\def\Bit{{\rm Bit}}
\def\LK{{\rm LK}}
\def\Head{{\rm head}}
\def\Tail{{\rm tail}}


\def\Univ{{\rm Univ}}
\def\Left{{\rm Left}}
\def\LeftFin{{\tt LeftFin}}
\def\Right{{\rm Right}}
\def\ES{{\rm ES}}
\def\PC{{\rm PC}}
\def\Fin{{\rm Fin}}
\def\CLJIDomega{{{\bf CLJID}^\omega}}
\def\CLJIDHA{{{\bf CLJID}^\omega+{\bf HA}}}
\def\LJIDHA{{{\bf LJID}+{\bf HA}}}
\def\LJID{{\bf LJID}}
\def\OP{{\rm OP}}
\def\Div{{\rm Div}}
\def\RP{{\rm RP}}
\def\MS{{\rm MS}}
\def\ET{{\rm ET}}
\def\LiftedTree{{\rm LiftedTree}}
\def\First{{\rm first}}
\def\Er{{\rm Er}}
\def\ER{{\rm ER}}
\def\Insert{{\rm insert}}
\def\ET{{\rm ET}}
\def\SInd{{\rm SInd}}
\def\Trans{{\rm Trans}}
\def\HA{{\bf HA}}
\def\DM{{\rm DM}}
\def\Ext{{\rm ext}}
\def\KB{{\rm KB}}
\def\DT{{\rm DT}}
\def\DS{{\rm DS}}
\def\Monoseq{{\rm Monoseq}}
\def\Eq{{\rm seq}}
\def\Last{{\rm last}}


\def\Uor{\biguplus}
\def\CLKIDomega{{{\bf CLKID}^\omega}}
\def\CLKIDPA{{{\bf CLKID}^\omega+{\bf PA}}}
\def\LKIDPA{{{\bf LKID}+{\bf PA}}}
\def\PA{{\bf PA}}


\def\LKIDN{{\bf LKIDN}}
\def\CLKIDN{{\bf CLKIDN}}


\def\Choose{{\rm Choose}}
\def\Asym{{\rm Asym}}
\def\Peano{{\bf Peano}}
\def\Code#1{{\lceil{#1}\rceil}}
\def\Ineq{{\rm Ineq}}
\def\MonoPath{{\rm MonoPath}}
\def\BoundedPath{{\rm BoundedPath}}
\def\InfPath{{\rm InfPath}}
\def\InfMonoPath{{\rm InfMonoPath}}
\def\HomSeq{{\rm HomSeq}}
\def\InfHomSeq{{\rm InfHomSeq}}

\def\ErdosTree{{\rm ErdosTree}}
\def\KB{{\rm KB}}
\def\MonoList{{\rm MonoList}}
\def\Univ{{\rm univ}}
\def\Infinite{{\rm Infinite}}
\def\S{{\bf S}}
\def\Ind{{\rm Ind}}
\def\LKID{{\bf LKID}}
\def\PA{{\bf PA}}
\def\LKIDExt{{\bf LKIDExt}}
\def\CLKID{{\bf CLKID^\omega}}


\def\Over#1#2{\deduce{#2}{#1}}
\def\List{{\rm List}}
\def\ListX{{\rm ListX}}
\def\ListO{{\rm ListO}}
\def\ListE{{\rm ListE}}
\def\Nl{{\rm nl}}


\def\SLLID{{\rm SLID}}
\def\SLID{{\rm SLID}}
\def\CSLID{{\rm CSLID}}
\def\CSLIDomega{{{\rm CSLID}\omega}}
\def\Eclass{{\rm Eclass}}
\def\Eq{{\rm Eq}}
\def\Deq{{\rm Deq}}
\def\Satom{{\rm Satom}}
\def\Cells{{\rm Cells}}
\def\Roots{{\rm Roots}}
\def\Elim{{\rm Elim}}
\def\Case{{\rm Case}}
\def\Unfold{{\rm Unfold}}
\def\Pred{{\rm Pred}}
\def\Start{{\rm Start}}
\def\Unsat{{\rm Unsat}}
\def\Split{{\rm Split}}
\def\Underscore{\underline{\phantom{x}}}
\def\Rootcell{{\rm rootcell}}
\def\Rootshape{{\rm rootshape}}
\def\Jointleaf{{\rm jointleaf}}
\def\Jointleafimplicit{{\rm jointleafimplicit}}
\def\Jointnode{{\rm jointnode}}
\def\Directjoint{{\rm directjoint}}


\def\Range{{\rm Range}}
\def\Xstart{X_{{\rm start}}}
\def\kstart{k_{{\rm start}}}
\def\Leaves{{\rm Leaves}}
\def\PureDist{{\rm PureDist}}
\def\Pure{{\rm Pure}}
\def\Identity{{\rm Identity}}
\def\Subst{{\rm Subst}}


\def\Connected{{\rm Connected}}
\def\Eststablished{{\rm Eststablished}}
\def\Valued{{\rm Valued}}


\def\SLRDbtw{{\rm SLRD}_{btw}}
\def\BTW{{{\rm BTW}}}
\def\Loc{{{\rm Loc}}}
\def\Ne{{\ne}}
\def\Nil{{{\rm nil}}}
\def\eqDef{=_{{\rm def}}}
\def\Inf#1{{\infty_{#1}}}
\def\Stores{{\rm Stores}}
\def\SVars{{{\rm SVars}}}
\def\Val{{{\rm Val}}}
\def\MSO{{{\rm MSO}}}
\def\Sep{{{\rm Sep}}}
\def\Septwo{{{\rm SLMI}}}
\def\SLMI{{{\rm SLMI}}}
\def\Sepinf{{\rm Sep}\infty}
\def\THeaps{{\rm THeaps}}
\def\Root{{\rm Root}}
\def\TG{{\rm TGraph}}


\def\Nil{{{\rm nil}}}
\def\eqDef{=_{{\rm def}}}
\def\Inf#1{{\infty_{#1}}}
\def\Stores{{\rm Stores}}
\def\SVars{{{\rm SVars}}}
\def\Val{{{\rm Val}}}
\def\MSO{{{\rm MSO}}}
\def\Sep{{{\rm sep}}}
\def\THeaps{{\rm THeaps}}
\def\Root{{\rm Root}}
\def\TG{{\rm TGraph}}

\def\Tilde{\widetilde}
\def\Bar{\overline}
\def\Lequiv{\Longleftrightarrow}
\def\Lto{\Longrightarrow}
\def\Lfrom{\Longleftarrow}
\def\Norm{{\rm Norm}}
\def\Noshare{{\rm Noshare}}
\def\Roots{{\rm Roots}}
\def\Forest{{\rm Forest}}
\def\Var{{\sf Var}}
\def\Range{{\rm Range}}
\def\Cell{{\rm Cell}}
\def\Switch{{\rm Switch}}
\def\All{{\rm All}}
\def\To{\leadsto}
\def\tree{{\tt tree}}
\def\Paths{{\rm Paths}}
\def\Finite{{\rm Finite}}
\def\Const{{\rm Const}}
\def\Leaf{{\rm Leaf}}
\def\LeastElem{{\rm LeastElem}}
\def\LeastIndex{{\rm LeastIndex}}
\def\WSnS{{{\rm WSnS}}}
\def\Expand{{{\rm Expand}}}


\long\def\J#1{} 
\def\T#1{\hbox{\color{green}{$\clubsuit #1$}}}
\def\W#1{\hbox{\color{Orange}{$\spadesuit #1$}}}

\def\Node{{{\rm Node}}}
\def\LL{{{\rm LL}}}
\def\DSN{{{\rm DSN}}}
\def\DCL{{{\rm DCL}}}
\def\LS{{{\rm LS}}}
\def\Ls{{{\rm ls}}}

\def\FPV{{{\rm FPV}}}
\def\Lfp{{{\rm lfp}}}
\def\IsHeap{{{\rm IsHeap}}}

\def\Equiv{\quad \equiv\quad }
\def\Null{{{\rm null}}}
\def\Emp{{{\rm emp}}}
\def\If{{{\rm if\ }}}
\def\Then{{{\rm \ then\ }}}
\def\Else{{{\rm \ else\ }}}
\def\While{{{\rm while\ }}}
\def\Do{{{\rm \ do\ }}}
\def\Cons{{{\rm cons}}}
\def\Dispose{{{\rm dispose}}}
\def\Vars{{{\rm Vars}}}
\def\Locs{{{\rm Locs}}}
\def\States{{{\rm States}}}
\def\Heaps{{{\rm Heaps}}}
\def\FV{{{\rm FV}}}
\def\BV{{{\rm BV}}}
\def\True{{{\rm true}}}
\def\False{{{\rm false}}}
\def\Dom{{{\rm Dom}}}
\def\dom{{{\rm dom}}}
\def\Abort{{{\rm abort}}}
\def\New{{{\rm New}}}
\def\W{{{\rm W}}}
\def\Pair{{{\rm Pair}}}
\def\Lh{{{\rm Lh}}}
\def\lh{{{\rm lh}}}
\def\Elem{{{\rm Elem}}}
\def\EEval{{{\rm EEval}}}
\def\PEval{{{\rm PEval}}}
\def\HEval{{{\rm HEval}}}
\def\EVal{{{\rm Eval}}}
\def\Domain{{{\rm Domain}}}
\def\Exec{{{\rm Exec}}}
\def\Store{{{\rm Store}}}
\def\Heap{{{\rm Heap}}}
\def\Storecode{{{\rm Storecode}}}
\def\Heapcode{{{\rm Heapcode}}}
\def\Lesslh{{{\rm Lesslh}}}
\def\Addseq{{{\rm Addseq}}}
\def\Separate{{{\rm Separate}}}
\def\Result{{{\rm Result}}}
\def\Lookup{{{\rm Lookup}}}
\def\ChangeStore{{{\rm ChangeStore}}}
\def\ChangeHeap{{{\rm ChangeHeap}}}
\def\Wand{\mathbin{\hbox{\hbox{---}$*$}}}
\def\Eval#1{\llbracket{#1}\rrbracket}
\def\Vec{\overrightarrow}

\def\Tilde{\widetilde}
\def\Break{\hfil\break\hbox{}}

\newcommand{\Set}[1]{\{ #1 \}}

\newcommand{\ap}                { {\tt ap} }
\newcommand{\cond}            { {\tt cond} }
\newcommand{\mapsfrom} {\mathbin{\leftarrow\!\!\rule[0.3pt]{0.5pt}{4pt}\,}}
\newcommand{\reduces}       { \mathbin{\mapsto} }
\newcommand{\CTlambda}   {{ {\sf CT}\mbox{-}\lambda }}
\newcommand{\RTlambda}   {{ {\sf  Reg}       \mbox{-}\Lambda^{\infty}_T }}

\newcommand{\RowTerm}    {{ {\sf Row}       \mbox{-}\Lambda^{\infty}_T }}
\newcommand{\FFVTerm}    {{ {\sf FinVar}    \mbox{-}\Lambda^{\infty}_T }}
\newcommand{\WTlambda}  {{ {\sf WTyped} \mbox{-}\Lambda^{\infty}_T }}
\newcommand{\GTC}            {{ {\sf GTC}       \mbox{-}\Lambda^{\infty}_T }}

\newcommand{\CT}                { {\tt CT} }
\newcommand{\etaRule}       { {\eta\mbox{-rule}} }
\newcommand{\ins}               { {\tt ins} }
\newcommand{\struct}          { {\tt struct} }
\newcommand{\Iter}              { {\tt It} }
\newcommand{\iter}               { {\tt it} }
\newcommand{\Interval}       { {\tt Iv} }
\newcommand{\interval}        { {\tt iv} }
\newcommand{\Succ}             { {\tt succ} }
\newcommand{\Zero}	    {{\tt 0}}
\newcommand{\nil}                { {\tt nil} }
\newcommand{\cons}             { {\tt c} }
\newcommand{\outline}  [2]   { {\contour{black}{\textcolor{#2}{#1}}}  }
\newcommand{\systemT}       { {\outline{\tt T}{white} } }
\newcommand{\LAMBDA}       { {  \outline{$\Lambda$}{white} } }
\newcommand{\Rec}               { {\tt Rec} }
\newcommand{\rec}                { {\tt rec} }
\newcommand{\exch}             { {\tt exch} }
\newcommand{\composed}     { {\mbox{\tiny $\circ$}} }
\newcommand{\WTyped}        { {\sf WTyped} }
\newcommand{\safe}              { {\hbox{\scriptsize \tt safe}} }
\newcommand{\GlobalTraceCondition}  { {\tt GlTrCon} }
\newcommand{\Sum}              { {\tt sum} }
\newcommand{\safeReduces} {\mathbin{\reduces_\safe}}
\newcommand{\nsafeReduces}[1] {\mathbin{\reduces_{#1\text{-}\safe}}}
\newcommand{\nsafeReducesL}[1] {\mathbin{\mapsfrom_{#1\text{-}\safe}}}
\newcommand{\nsafeReducesAstL}[1] {\mathbin{\mapsfrom_{#1\text{-}\safe}^*}}
\newcommand{\safeReducesAst} {\mathbin{\reduces_\safe^*}}
\newcommand{\safeReducesPlus} {\mathbin{\reduces_\safe^+}}
\newcommand{\nsafeReducesAst}[1] {\mathbin{\reduces_{#1\text{-}\safe}^*}}
\newcommand{\nsafePReduces}[1] {\mathbin{\Rightarrow_{#1}}}
\newcommand{\nsafePReducesL}[1] {\mathbin{\Leftarrow_{#1}}}
\newcommand{\nsafePReducesMany}[2] {\mathbin{\Rightarrow_{#1}^{#2}}}
\newcommand{\nsafePReducesManyL}[2] {\mathbin{\Leftarrow_{#1}^{#2}}}
\newcommand{\nequal}[1] {\mathbin{\approx_{#1}}}
\newcommand{\var}                { {\tt var} }
\newcommand{\Graph}            { {\tt Graph} }
\newcommand{\bfColor} [2]     {{\bf \color{#1}{#2}}}
\newcommand{\weak}             {{\tt wk}}
\newcommand{\RED}{{\textsf{RED}}}

\definecolor{amber}{rgb}{1.0, 0.75, 0.0}
\definecolor{oldgold}{rgb}{0.81, 0.71, 0.23}

\newcommand{\SubTerm}               {{\rm SubT}}
\newcommand{\LSubTerm}[2]           {{\tt Sub_{#1}(#2)}}
\newcommand{\Lctx}[1]           {{\tt C}_{#1}}
\newcommand{\Lv}[2]           {||#2||_{#1}}
\newcommand{\Tree}                      {{\tt Tree}}
\newcommand{\N}                           {{N}}
\newcommand{\Num}                      {{\tt Num}}
\newcommand{\Nat}                        {\mathbb{N}}
\newcommand{\Reg}                       {{\tt Reg}}
\newcommand{\Type}                      {{\sf Type}}
\newcommand{\conc}                      {{\star}}
\newcommand{\typingRule}            {{\tt r}}
\newcommand{\apvar}                    {{{\tt ap}_{\mbox{\tiny $v$}}}}
\newcommand{\apnotvar}               {{{\tt ap}_{\mbox{\tiny $\neg v$}}}}
\newcommand{\subseteqsim}         {\mathbin{\raisebox{2pt}{$\subset$}_{\!\!\!\!\!\sim}}}
\newcommand{\supseteqsim}         {\mathbin{\raisebox{2pt}{$\supset$}_{\!\!\!\!\!\sim}}}
\newcommand{\Ctxt}                       {{\tt Ctxt}}
\newcommand{\Seq}                        {{\tt Seq}}
\newcommand{\Judg}                       {{\tt Judg}}
\newcommand{\Rule}                       {{\tt Rule}}
\newcommand{\restr}                      {{\mid}} 
\newcommand{\setT}                       {{\mathcal T}}
\newcommand{\setM}                       {{\mathcal M}}
\newcommand{\base}                      {{\tt b}}
\newcommand{\inductive}               {{\tt ind}}
\newcommand{\universe}     [1]       {{|#1|}}
\newcommand{\Label}                     {{\tt Lbl}}
\newcommand{\quotationMarks} [1]     {{\textquotedblleft#1\textquotedblright}}
\newcommand{\lexicographic} [1]   {{ <^{#1}_{\mbox{\tiny lex}}} }
\newcommand{\trunk}                     {{\tt trunk}}
\newcommand{\Atom}{{\tt at}}
\newcommand{\ls}{\ell}

\newcommand\Restrict[2]{#1_{#2}}
\newcommand\simIndex[1]{\sim^{#1}_{{\tt idx}}}
\newcommand\mergeCtx[2]{#1\sharp#2}
\newcommand\Ifz{{\tt ifz}}
\newcommand\TtoCT[1]{(#1)^\circ}

\definecolor{purple}{rgb}{0.8, 0.2, 0.8}

\newcommand{\Rapv}{\hbox{$(\text{ap}_{\text{v}})$}}
\newcommand{\RapNv}{\hbox{$(\text{ap}_{\lnot\text{v}})$}}
\newcommand{\Reta}{\hbox{$(\eta)$}}
\newcommand{\Rap}{\hbox{$(\text{ap})$}}
\newcommand{\Rcond}{\hbox{$(\text{cond})$}}
\newcommand{\Ack}{{\tt Ack}}
\newcommand{\AckA}{{\tt ack1}}
\newcommand{\AckB}{{\tt ack2}}
\newcommand{\ACK}{{\redM Ack}}
\newcommand{\Cond}[2]{{\tt cond}(#1,#2)}
\newcommand{\Suc}[1]{{\tt S}(#1)}
\newcommand{\goldN}{\bfColor{oldgold}{\mathit{N}}}
\newcommand{\goldNout}{{\contour{black}{{\color{oldgold}{{\it N}}}}}}
\newcommand{\golddagger}{{\contour{black}{{\color{oldgold}{$\dagger$}}}}}
\newcommand{\goldspadesuit} 
      {{ \bf \contour{black}{\textcolor{oldgold}{$\spadesuit$}} }}
\newcommand{\redN}{{\bfColor{red}{\mathit{N}}}}
\newcommand{\bluen}{{\bfColor{blue}{n}}}
\newcommand{\redm}{{\bfColor{red}{m}}}
\newcommand{\blueN}{{\bfColor{blue}{\mathit{N}}}}
\newcommand{\orangeN}{{\bfColor{orange}{\mathit{N}}}}
\newcommand{\redblueN}{{\bfColor{red}{\mathit{N}}\hspace{-1em}\bfColor{blue}{\mathit{N}}}}

\newcommand{\Pfin}                       {{\setP_{\mbox{\tiny fin}}}}
\newcommand{\eqMap}                  {{\tt eq}}
\newcommand{\BadSeq}                 {{\tt BadS}}

\newcommand{\labelled}                  { {\mbox{\tiny \tt lab}} }  
\newcommand{\CTlambdaLabelled} {{\mbox{\tt lab-}\CTlambda}}
\newcommand{\parallelReduces}         { {  \mapsto \mspace{-17mu} \mapsto  }}
\newcommand{\Lim}                            {{\tt Lim}}
\newcommand{\length}                     {{\tt len}}
\newcommand{\form}                     {{\tt form}}
\def\head                                         {{\rm head}}   
                  
\newcommand{\Par}{\,\|\,}
\newcommand{\trunc}[2]{{#1}^{[#2]}}
\newcommand{\dist}{\textit{d}}
\newcommand{\infixapp}{\MVAt}
\renewcommand{\Subst}[3]{#1[#2 := #3]}
\newcommand{\tuple}[1]{\langle #1 \rangle}
\newcommand{\der}{\vdash}
\newcommand{\sqleq}{\sqsubseteq}
\newcommand{\Snum}[1]{\overline{#1}}
\newcommand{\trans}[1]{{\sharp#1}}
\newcommand{\infreduces}{\mathbin{\mapsto}^\infty}
\newcommand{\depth} {{\tt depth}}
\newcommand{\height} {{\tt h}}

\newcommand{\Arg}								{{\sf Arg}}
\newcommand{\Restr}						   {{\sf Restr}}
\newcommand{\tail}                        {{\sf tail}}
\newcommand{\argPair} [2]            {{(#1 | #2)}}
\newcommand{\Path}                      {{\sf Path}}
\newcommand{\GTCrestr}              {{\mbox{GTC}_{\tiny \sf restr}}}
\newcommand{\infinitaryT}            {{\outline{\tt T}{black}_{\infty}}}
\newcommand{\restrT}                    {{\outline{\tt T}{black}_{\tiny \sf restr}}}
\newcommand{\restrC}                    {{\mu}}

\section{Infinitary Lambda Terms}\label{sec:terms}
We extend the infinitary $\lambda$-calculus $\Lambda^\infty$
\cite{KENNAWAY199793,KennawayEtAlii2003}.

of Kennaway et alii, 
by adding the constant $0$ and the function symbols $\Succ$ and $\cond$. Then
we define the subset $\GTC$ of terms satisfying the Global Trace Condition (GTC), and for
them we prove that all reductions are strongly convergent and all terms are total. $\GTC$ is an extension with variables and $\lambda$ of {\em CT},  Cyclic System $T$ of Das \cite{2021-Anupam-Das,DBLP:conf/fscd/000221}. 

Our definition of $\GTC$ is by consecutive restrictions. 
We first define the set $\RowTerm$ of infinitary row-terms with $\lambda$ and conditional, most of which are meaningless, for which neither $\alpha$-renaming nor substitution is sound. Then we define the subset $\FFVTerm$ of row-terms with finitely many free variables, for which we can consistently define $\alpha$-renaming and substitution. Then we define the subset $\WTyped$ of infinitary terms with finitely many variables that are well-typed. Eventually, we restrict well-typed infinitary terms by adding GTC, a sufficient condition for totality.

\begin{definition}[Types and the set $\RowTerm$ of Row-Terms]\label{def:Types-RowTerms}
Let $\Type$ be the set inductively defined by
\[\Type \ni A,B ::= N \mid (A\to B)\]
Then the set $\RowTerm$ is coinductively defined by
\[\RowTerm \ni t,s ::= x \mid (\lambda x{:}A. t) \mid (t\,s) \mid 0 \mid \Succ(t) \mid \cond(t,s)\]
where $x$ ranges over a denumerable set $\Var$ of term variables.
\end{definition}

As usual, we omit unnecessary braces, assuming that $\to$ associates to the right and term application associates to the left. 
We abbreviate by $A_1, \ldots , A_n \to B$, and by $\vec{A} \to B$ when the $A_i$ are understood, the type $(A_1 \to \cdots (A_n \to B)\cdots)$, and say that the $A_i$ are in {\em argument position}.

$\Type$ is the set of simple types generated by the single atomic type $N$, whose intended interpretation is the set of natural numbers. Row-terms are possibly infinite terms representing elements of the metric completion of finite ones, namely those with a finite syntactic tree. Unlike \cite{KENNAWAY199793} and following \cite{KurzPSV12}, we consider the subset $\FFVTerm$ of raw terms with finite $\FV(t)$, the set of free variables occurring in $t$, which admits the obvious corecursive definition. We also adapt the slightly different metric from \cite{KurzPSV12} to account for truncation.

\begin{definition}\label{def:cut-metric}
Let $\FFVTerm + \bot$ be $\FFVTerm$ extended with the constant $\bot$. 
Then for all $n \in \Nat$ we define the map
$\trunc{(-)}{n}: \FFVTerm \to \FFVTerm + \bot$ by $\trunc{t}{0} = \bot$ and
\[\begin{array}{rcl@{\hspace{0.8cm}}rcl@{\hspace{0.8cm}}rcl}
\trunc{x}{n+1} & = & x & \trunc{(ts)}{n+1} & = & \trunc{t}{n} \trunc{s}{n} & 
                     \trunc{(\lambda x{:}A.t)}{n+1} & = & \lambda x{:}A.\trunc{t}{n} \\ [1mm]
\trunc{\Zero}{n+1} & = & \Zero & \trunc{\Succ(t)}{n+1} & = & \Succ(\trunc{t}{n})
                     & \trunc{\cond(t,s)}{n+1} & = & \cond(\trunc{t}{n}, \trunc{s}{n})
\end{array}\]
The metric $\dist: (\FFVTerm + \bot)^2 \to [0, 1]$ is $\dist(t,s) = 2^{-m}$ where
$m$ is the last $n \in \Nat$ such that $\trunc{t}{n} =  \trunc{s}{n}$ if it exists, $\dist(t,s) = 0$
otherwise.
\end{definition}

The choice of $\FFVTerm$ makes it easier and unproblematic to define $\alpha$-equivalence and substitution, which are as follows.

\begin{definition}[$\alpha$-Equivalence on $\FFVTerm$]
\label{def:alpha-equivalence}
Let $=_\alpha$ be the ordinary $\alpha$-equivalence over finite terms of system T + $\bot$. For any $t, u \in \FFVTerm$ define \[t =^\infty_\alpha u \qquad \mbox{iff} \qquad \forall n \in \Nat.\; \trunc{t}{n} =_\alpha \trunc{u}{n}\]
\end{definition}

\begin{proposition}\label{prop:alpha-equivalence-FFVTerm}
The relation $=^\infty_\alpha$ is a congruence over $\FFVTerm$.
\end{proposition}

In view of the above, we consider the terms in $\FFVTerm$ up to $\alpha$-equivalence, and define substitution corecursively as follows.

\begin{definition}[Substitution on  $\FFVTerm$]\label{def:substitution}
For $t,s \in \FFVTerm$ and $x \in \Var$ the {\rm substitution} of $s$ for $x$ in $t$, written $\Subst{t}{x}{s}$, is corecursively defined by:
\begin{enumerate}
\item $\Subst{x}{x}{s} = s$, and $\Subst{y}{x}{s} = y$ 
for $\Var \ni y \neq x$
\item $\Subst{(t\, t')}{x}{s} = (\Subst{t}{x}{s})(\Subst{t'}{x}{s})$
\item \label{case:abs} $\Subst{(\lambda y{:}A.t)}{x}{s} = \lambda z{:}A.\Subst{(\Subst{t}{y}{z})}{x}{s}$ for the first $z \not\in \Set{y} \cup \FV(t) \cup \FV(s)$
\item $\Subst{0}{x}{s} =0$ and $\Subst{\Succ(t)}{x}{s} = \Succ(\Subst{t}{x}{s})$
\item $\Subst{\cond(t,t')}{x}{s} = \cond(\Subst{t}{x}{s},\Subst{t'}{x}{s})$
\end{enumerate}
\end{definition}

\noindent
In the clause \ref{case:abs} above, the required $z$ always exists because $\FV(t) \cup \FV(s)$ 
is finite, and its choice is immaterial by $\alpha$-equivalence. 

Finally, a row term is {\em regular} if its syntactic tree is regular. We call $\RTlambda$ the set of regular terms, which is clearly a (proper) subset of $\FFVTerm$.

\medskip
Reduction rules for $\FFVTerm$, also called notions of reduction in the literature, include the usual $\beta$-reduction and an additional rule for the conditional.

\begin{definition}[Reduction rules]
  Let $f$, $t$ and $u$ be terms in $\FFVTerm$. We define the $\beta$-reduction relation $\reduces_\beta$
  and the $\cond$-reduction relation $\reduces_\cond$ as follows:
\begin{description}
\item[$\beta$-reduction:] $(\lambda x{:}A.t)(u) \reduces_\beta \Subst{t}{x}{u}$, 
\item[$\cond$-reduction:] $\cond(u,f)(0) \reduces_\cond u$ and $\cond(u,f)(\Succ (t)) \reduces_\cond f(t)$.
\end{description}
\end{definition}

As usual, the relation $\reduces$ is the compatible closure of $\reduces_\beta \, \cup \, \reduces_\cond$, and $\reduces^*$ is its reflexive and transitive closure.



\section{Well-typed terms and Global Trace Condition}\label{sec:typing.system}

\begin{figure}
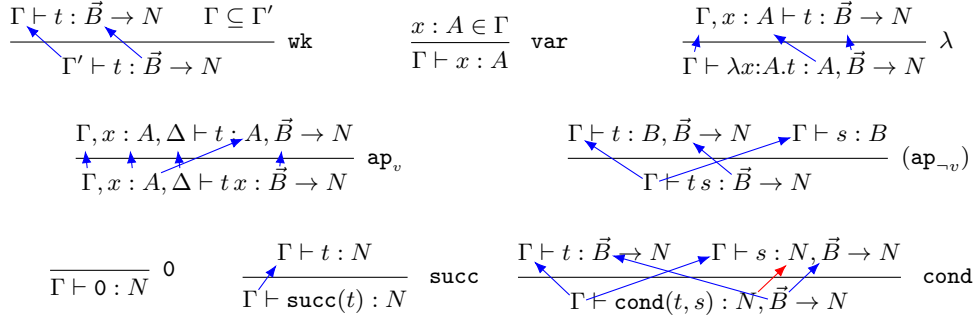

	
	\begin{tabular}{ccc}
	
		\begin{minipage}{0.33\textwidth}
		    	\centering
			    \begin{prooftree}
			      \hypo{\vnode1{\Gamma} \vdash t: \vnode2{\vec{B}} \rightarrow N}
			      \hypo{\Gamma \subseteq \Gamma'}
			      \infer2[\weak]{\vnode3{\Gamma'} \vdash t: \vnode4{\vec{B}} \rightarrow N} 
			    \end{prooftree}
			    $
			    \dirflowblueedges{node3/node1}
			    \dirflowblueedges{node4/node2} 
			    $
		 \end{minipage}
	  	
	  	&

		\begin{minipage}{0.23\textwidth}
	    	\centering
		    \begin{prooftree}
		      \hypo{x:A \in \Gamma}
		      \infer1[\var]{\Gamma \der x:A} 
		    \end{prooftree}

	  	\end{minipage}
	  	
	  	&

	  	\begin{minipage}{0.33\textwidth}
		    	\centering
			    \begin{prooftree}
			      \hypo{\vnode1{\Gamma}, \vnode2{x:A} \vdash t : \vnode3{\vec{B}} \rightarrow N}
			      \infer1[$\lambda$]{\vnode4{\Gamma} \vdash \lambda x{:}A.t: \vnode5{A}, \vnode6{\vec{B}} \rightarrow N} 
			    \end{prooftree}
			    $
			    \dirflowblueedges{node4/node1}
			    \dirflowblueedges{node5/node2}
			    \dirflowblueedges{node6/node3} 
			    $
	  	\end{minipage}
	  	
		\bigskip
		
		\end{tabular}
		
		\begin{tabular}{cc}
		
  			\begin{minipage}{0.47\textwidth}
		    	\centering
			    \begin{prooftree}
			      \hypo{\vnode1{\Gamma}, \vnode8{x:A}, \vnode9{\Delta} 
			      \vdash t : \vnode2{A}, \vnode3{\vec{B}} \rightarrow N}
			      \infer1[$\apvar$]{\vnode4{\Gamma},\vnode6{x:A}, \vnode7{\Delta} 
			      \vdash t \,x: \vnode5{\vec{B}} \rightarrow N} 
			    \end{prooftree}

			    $
			    \dirflowblueedges{node4/node1}
			    \dirflowblueedges{node6/node2}
			    \dirflowblueedges{node6/node8}
			    \dirflowblueedges{node5/node3} 
			    \dirflowblueedges{node7/node9}
			    $
		  	\end{minipage}

	  		&

	  		\begin{minipage}{0.47\textwidth}
		    	\centering
			    \begin{prooftree}
			      \hypo{\vnode1{\Gamma} \vdash t : B, \vnode2{\vec{B}} \rightarrow N}
			      \hypo{\vnode3{\Gamma} \vdash s : B}
			      \infer2[($\ap_{\neg v}$)]{\vnode4{\Gamma} \vdash t \, s: \vnode5{\vec{B}} \rightarrow N} 
			    \end{prooftree}

			    $
			    \dirflowblueedges{node4/node1}
			    \dirflowblueedges{node4/node3}
			    \dirflowblueedges{node5/node2} 
			    $
		  	\end{minipage}

		\end{tabular}
		
		\bigskip
		
		\begin{tabular}{ccc}
		
		\begin{minipage}{0.23\textwidth}
	    	\centering
		    \begin{prooftree}
		      \hypo{}
		      \infer1[\Zero]{\Gamma \vdash \Zero: N} 
		    \end{prooftree}

	  	\end{minipage}
		
		\begin{minipage}{0.23\textwidth}
	    	\centering
		    \begin{prooftree}
		      \hypo{\vnode1{\Gamma} \vdash t: N}
		      \infer1[\Succ]{\vnode2{\Gamma} \vdash \Succ(t): N} 
		    \end{prooftree}

		    $
		    \dirflowblueedges{node2/node1} 
		    $
	  	\end{minipage}
		
		&  
		  	\begin{minipage}{0.33\textwidth}
		    	\centering
			    \begin{prooftree}
			      \hypo{\vnode1{\Gamma} \vdash t : \vnode2{\vec{B}} \rightarrow N}
			      \hypo{\vnode3{\Gamma} \vdash s : \vnode4{N}, \vnode5{\vec{B}} \rightarrow N}
			      \infer2[\cond]{\vnode6{\Gamma} \vdash \cond(t,s) : \vnode7{N}, \vnode8{\vec{B}} \rightarrow N} 
			    \end{prooftree}

			    $
			    \dirflowblueedges{node6/node1}
			    \dirflowblueedges{node6/node3}
			    \dirflowrededges{node7/node4}
			    \dirflowblueedges{node8/node2}
			    \dirflowblueedges{node8/node5} 
			    $

		  	\end{minipage}
		
		\end{tabular}

	\caption{Typing rules and the trace relation}
   	\label{fig:WST-rulesTraces}

\end{figure}

To rule out meaningless row-terms like $0\,0$ or $\Succ(\lambda x{:}N.x)$, we introduce the typing system in Figure \ref{fig:WST-rulesTraces} for terms in $\FFVTerm$. 
There are differences between our system and the inference rules of the simply typed $\lambda$-calculus and G\"odel System $T$. In our system, contexts are sequences of variable typings rather than sets. By $x:A \in \Gamma$, we mean that $x:A$ occurs in $\Gamma$, and $\Gamma \subseteq \Gamma'$ means that there is an injection from the first sequence into the second, possibly including permutations. We explicitly list the rule $\weak$ for weakening+exchange, which is not merely admissible. 
Further, the application rule is split into the rules $\apvar$ and
$\ap_{\neg v}$, depending on whether the term $s$ in the subject  $t\,s$ of the conclusion is a variable: $\apvar$ corresponds to a contraction between two occurrences of $A$ in Logic. Finally, we add a pair of relations, represented by {\bf \color{blue}{blue}} and {\bf \color{red}{red}} arrows, linking variables in type contexts and types in the argument position of the derived type in the conclusion to variables and types in the premises of each rule. These relations are essential to the definition of traces below.

\begin{definition}[Derivation tree]
\begin{enumerate}
\item A {\em judgment} is a triple $\Gamma \der t:A$, where $\Gamma = x_1:A, \ldots, x_n : A_n$ is a {\em context}, namely a finite sequence of pairs $x_i:A_i$
	with $x_i \in \Var$ and $A_i \in \Type$ and with pairwise distinct term variables, $t \in \FFVTerm$ called the subject, and $A \in \Type$ called the derived type.
\item A {\em derivation} is any finite or infinite tree $\Pi$ whose nodes are labeled with judgments and whose edges are instances of the rules in 
        Figure \ref{fig:WST-rulesTraces}, without infinite nesting of the $\weak$-rule. 
\item $\Gamma \vdash t :A$ is {\em derivable} if there exists a derivation $\Pi$ with root $\Gamma \vdash t :A$; then we say that $t$ is {\em well-typed}, and denote
	by $\WTyped$ the set of such terms.
\end{enumerate}
\end{definition}

We observe that both terms and derivations may be infinite trees, and only finite terms in $\WTyped$ are typable by finite derivations.
A well-typed term does not necessarily have a unique derivation and type, even when the type context is fixed.
Consider the family of closed terms $t_n = t_{n+1} \,0$; 
then for each $n \in \mathds{N}$ we can find a derivation of $\vdash t_n :N^n \rightarrow A$, where $N^n$ is a list of $n$ occurrences of $N$ and $A$ is arbitrary. 
Moreover, derivations are too weak to filter out infinite terms of the form $t = (\lambda x{:}N.t)\,0: A$ or $t = \cond(t, \lambda x{:}N.t)\,\Succ(0): A$ that reduce infinitely often to a redex, never returning a meaningful output. These terms are called \emph{active} in Kennaway et alii \cite{KENNAWAY199793,KennawayEtAlii2003}. We introduce the {\em Global Trace Condition} (GTC) for our system by adapting concepts from \cite{BrotherstonSimpson2011,DBLP:conf/fscd/000221} that rule out active well-typed terms. Moreover, the condition GTC rules out closed terms of type $N$ that do not denote numbers, such as $t = \Succ(t) : N$, or do not denote total functionals, and enforces the uniqueness of typing.

\begin{definition}[Traces and GTC]
\begin{enumerate}
\item  For each rule $r$ in Figure \ref{fig:WST-rulesTraces}, variables in the contexts and types in argument position of the derived types, which are connected by either a  
          {\color{blue}{blue}} arrow, called stationary, or a {\color{red}{red}} arrow, called progressing, in Figure \ref{fig:WST-rulesTraces}, are in {\em trace relation}.
          With $ \Gamma \,{\color{blue}{\to}}\, \Gamma'$ we abbreviate the trace relation $x:A \, {\color{blue}{\to}} \, x:A$ for all $x:A$ both in $\Gamma$ and $\Gamma'$.

\item For any derivation $\Pi$ of $\Gamma \vdash t:A$, and branch $\pi$ of $\Pi$, a trace following $\pi$ is any chain of trace relations among nodes of $\pi$. 
 
\item A {\em progress point} of a trace is a trace relation colored with {\color{red}{red}}, between the first argument $N$ of $\cond(t,s): N,\vec{B} \rightarrow N$ 
  	and the first argument $N$ of $s: N,\vec{B} \rightarrow N$ in the right premise of the rule. A trace is {\em progressing} if it includes a progress point.
\item A derivation $\Pi$ satisfies the GTC if for all infinite branches $\pi$ of $\Pi$, there is some trace from a node of $\pi$ progressing infinitely many times. 
	Such a derivation is called a proof.
  
\item We denote by $\GTC$ the set of terms in $\WTyped$ that have a proof.
\end{enumerate}
\end{definition}

In Figure \ref{fig:sum-proof}, we illustrate the proof of the typing of the term $\Sum = \lambda x{:}\N. \cond(x,\lambda y{:}\N.\Succ(\Sum\,x\,y))$ for summation.
We depict only the relevant arrows of the trace relations. Since $\Sum \in \RTlambda$ (namely, it is regular), the proof itself is regular, and we represent it by a finite
cyclic graph, where the only cycle is represented by the 
pair of the two $(\dagger)$ connected by a black arrow.

\begin{figure}[t]
\begin{center}
    \begin{prooftree}
    	\hypo{}
    	\infer1[$\var$]{x:N \der x:N}
	\hypo{\der \Sum:\N \to \vnode8{\N} \to \N \qquad (\vnode9{\dagger}) }
	\infer1[$\weak$]{x:\N, y : \vnode7{\N} \der \Sum:\N \to \N \to \N}
	\infer1[$\apvar$]{x:\N, y : \vnode6{\N} \der \Sum\,x:\N \to \N}
	\infer1[$\apvar$]{x:\N, y : \vnode5{\N} \der \Sum\,x\,y:\N}
	\infer1[$\Succ$]{x:\N, y : \vnode4{\N} \der \Succ(\Sum\,x\,y): \N}
	\infer1[$\lambda$]{x:\N \der \lambda y{:}\N.\Succ(\Sum\,x\,y): \vnode3{\N} \to \N}
	\infer2[$\cond$]{x:N \der \cond(x,\lambda y{:}\N.\Succ(\Sum\,x\,y)): \vnode2{\N} \rightarrow \N}
	\infer1[$\lambda$]{\der \Sum:\N \rightarrow \vnode1{\N} \rightarrow \N \qquad (\vnode0{\dagger})}
    \end{prooftree}
    $
    \dirflowedges{node1/node2}
    \dirflowrededges{node2/node3}
  \dirflowedges{node3/node4}
  \dirflowedges{node4/node5}
  \dirflowedges{node5/node6}
  \dirflowedges{node6/node7}
  \dirflowedges{node7/node8}
  \bentdirflowedges{node8/node1/-90}
  \bentflowblackedges{node9/node0/120}
   $    
  \caption{Typing of $\Sum = \lambda x{:}\N. \cond(x,\lambda y{:}\N.\Succ(\Sum\,x\,y))$}\label{fig:sum-proof}
\end{center}
\end{figure}

\begin{theorem}[Uniqueness of Types and Subject Reduction]\label{thr:Uniqueness-SR}

Let $t \in \GTC$, then:
\begin{enumerate}
\item \label{theorem-uniqueness-typing}
For all $\Gamma$, there exists a unique type $A$ such that some $\Gamma \der t:A$ is provable. 
\item \label{thr:SR}
If there is a proof of $\Gamma \der t:A$ and $t \reduces t'$ then $\Gamma \der t':A$ is provable.
\end{enumerate}
\end{theorem}

The proof of $\Gamma \der t:A$ claimed in Theorem \ref{theorem-uniqueness-typing} is not unique in general. However, it can be transformed into a (unique) canonical form by:
\emph{(i)} restricting the conclusion $\Gamma' \der u:B$ of each instance of rule $\not = \weak$ so that $\dom(\Gamma') = \FV(u)$;
\emph{(ii)} collapsing any two consecutive $\weak$ into one; \emph{(iii)} removing identical $\weak$ and possibly adding at most one $\weak$ to the end. 

Intuitively, a progress point represents the application of the conditional $\cond(t, s)$ to decreasing values of its argument. All other references to argument terms or potential values of free variables are considered stationary, meaning their values in an actual computation need not decrease.
The GTC prevents infinite computations involving infinitely decreasing values of type $N$. This is stated in the next section, where we relate infinitely progressing traces to strongly convergent reductions of transfinite length.

\section{Infinite reduction and strong convergence}\label{sec:infinite-reductions}

Given an ordinal $\alpha$, either finite or infinite, a sequence of terms $(t_\beta)_{\beta < \alpha}$ is a {\em reduction of length $\alpha$} if for all $\beta < \alpha$ we have $t_\beta \reduces t_{\beta + 1}$.
In the theory of the infinitary $\lambda$-calculus, normal forms (terms not including any redex) might be reached only in the limit, after infinitely many reduction steps; consistently, the concept of normalizing reduction is replaced by that of {\em strongly convergent reduction}, which does not necessarily yield a normal form but rather generates a sequence of terms that is Cauchy-convergent with respect to the term distance and such that the depth of contracted redexes tends to infinity (see \cite{KennawayEtAlii2003}).

To reconcile strong convergence with the idea of strong normalization in type theory, we consider a stronger notion of possibly infinite reduction, which we call {\em $\cond$-convergence}. In the following, a position is a finite string $p \in \Set{1,2}^*$ representing a node in the syntactic tree of a term $t \in \FFVTerm$, where the empty sequence is the root of $t$; if $t = \lambda x{:}A.t'$ or $t = \Succ(t')$ and $p = 1\,q$, then $p$ is the position $q$ in $t'$; if $t = t't''$ or $t = \cond(t',t'')$, then $p = 1\,q$ is the position $q$ in $t'$, and $p = 2\,q$ is the position $q$ in $t''$. If $p$ is a position in $t$, then $t \upharpoonright p$ denotes the subterm of $t$ rooted at $p$.

\begin{definition}[$\cond$-Convergence]\label{def:cond-convergence}
A position $p$ in the syntactic tree of a term $t \in \FFVTerm$ has $\cond$-depth $n$ if the number of $2$'s in $p$ such that, for some $q$, $q\,2$ is a prefix of $p$ and $t \upharpoonright q$ is a $\cond$ term, is $n$.

A reduction  $(t_\beta)_{\beta < \alpha}$ is $\cond$-convergent if for all limit ordinals $\lambda \leq \alpha$,
the limit of the $\cond$-depth of the redex reduced in $t_\beta$ for $\beta < \lambda$ is infinite.
\end{definition}

In words, a position $p$ of a term $t$ encodes a finite path from the root of the syntactic tree of $t$; it has $\cond$-depth $n$ if it passes through the tree of the subterm $t_2$ of some occurrence of $\cond(t_1, t_2)$ in $t$ exactly $n$ times. Remember that $\cond(t_1,t_2)(\Succ(u)) \reduces t_2 (u)$, thereby mimicking the recursion mechanism via the nesting of $\cond$ in $t_2$. Hence, an infinite $\cond$-convergent reduction is one in which no redex on the right branch of a $\cond$ can be contracted infinitely often.

Any $\cond$-convergent reduction is strongly convergent, but not vice versa. We can establish the following results.

\begin{theorem}[Strong convergence of terms in $\GTC$]\label{thr:main}
Let $t \in \GTC$, then:
\begin{enumerate}
\item \label{thr:main-2}
	All reductions out of $t$ are $\cond$-convergent and, consequently, strongly convergent.
	\item \label{thr:main-3}
	All reductions from $t$ always reducing some redex of minimum $\cond$-depth end in at most $\omega$ steps to some normal form.  
	If, furthermore, $t$ is a closed term of type $N$, then such reductions terminate in finitely many steps in some $\Succ^n(0)$.
\end{enumerate}
\end{theorem}

We briefly mention additional properties of terms in $\GTC$ that are not covered in the paper. Typability is preserved not only under finite reductions, as stated in Theorem \ref{thr:Uniqueness-SR}.\ref{thr:SR}, but also in the limit.
As for $\Lambda^\infty$, confluence does not hold in general because of infinite reductions, but it holds in the limit. This allows us to show that the numeral in \ref{thr:main-3} above is unique. Finally, we conjecture that the expressive power of our system is the same as that of Das's system $CT$ and, hence, Goedel's system $T$.




\bibliography{cyclic}

\end{document}